\documentstyle[times,pramana,epsf,floats]{ias}

\def\pr{\prime}
\def\beq{\begin{eqnarray}}
\def\eeq{\end{eqnarray}}
\def\be{\begin{equation}}
\def\lan{\left\langle}
\def\ran{\right\rangle}
\def\ee{\end{equation}}
\def\barr{\begin{array}}
\def\earr{\end{array}}

\def\nn8{\\}
\def\l{\left}
\def\r{\right}
\def\dis{\displaystyle}
\def\ed{\end{document}}

\def\bF{{\mbox{\boldmath $F$}}}

\def\cq{{\cal Q}}
\def\caa{{\cal A}}

\def\cn{{\cal N}}

\def\cs{{\bf s}}
\def\ct{{\bf t}}

\def\spin{\frac{1}{2}}

\begin{document}

\mark{{EGUE(2)-$SU(4)$}{Manan Vyas and V. K. B. Kota}}
\title
{Random matrix ensembles with random interactions: Results
for EGUE(2)-$SU(4)$}

\author{Manan Vyas$^a$ and V. K. B. Kota$^{a,b}$}
\address{$^a$Physical Research Laboratory, Ahmedabad 380 009, India \\
$^b$Department of Physics, Laurentian University, Sudbury,
ON P3E 2C6, Canada}

\abstract{We introduce in this paper embedded Gaussian unitary ensemble of
random matrices, for $m$ fermions in $\Omega$ number of single particle
orbits, generated by random two-body interactions that are $SU(4)$
scalar, called EGUE(2)-$SU(4)$. Here the $SU(4)$ algebra corresponds to
Wigner's  supermultiplet  $SU(4)$ symmetry in nuclei.  Formulation based on
Wigner-Racah algebra of the embedding algebra $U(4\Omega) \supset U(\Omega)
\otimes SU(4)$ allows for analytical treatment of this ensemble and
using this analytical formulas are derived for the covariances in energy
centroids and spectral variances. It is found that these covariances
increase in magnitude as we go from EGUE(2) to EGUE(2)-$\cs$ to
EGUE(2)-$SU(4)$ implying that symmetries may be responsible for chaos in
finite interacting quantum systems.}

\maketitle

\section{Introduction}

Hamiltonians for finite quantum systems such as nuclei, atoms, quantum
dots, small metallic grains, interacting spin systems modeling quantum
computing core, Bose condensates and so on consist of interactions of low
body rank and therefore embedded Gaussian ensembles (EGE) of random
matrices generated by random interactions, first introduced in 1970 in the
context of nuclear shell model and explored to some extent in the 70's and
80's, are appropriate for these systems. Note that EGE's that correspond to
the classical ensembles GOE, GUE and GSE are EGOE, EGUE and EGSE
respectively. With the interest in many body chaos, EGE's received new
emphasis beginning from 1996 and since then a wide variety of EGE's have
been introduced in literature, both for fermion and boson systems
\cite{Ko-01,Pa-07,Ko-08,Ch-04}. See \cite{Ko-01,Pa-07,Ko-06a,Ma-09} and
references therein for recent applications of EGE's.

EGE's generated by two body interactions [EGE(2)] for spinless fermion
systems are the simplest of these ensembles. For $m$ fermions in $N$ single
particle (sp) states, the embedding algebra is $SU(N)$. It is well
established that $SU(N)$ Wigner-Racah algebra solves EGUE(2) and also the
more general EGUE($k$) as  well as EGOE($k$) \cite{Ko-05,Be-01}. Realistic
systems carry good quantum numbers (for example spin $S$ for quantum dots,
angular momentum $J$ for nuclei) in addition to particle number $m$,
therefore EGE's with good symmetries should be studied. EGUE(2)-$\cs$ and
EGOE(2)-$\cs$, for fermions with spin $\cs=\frac{1}{2}$ degree of freedom,
are the simplest non-trivial EGE's with immediate physical applications.
For $m$ fermions occupying $\Omega$ number of orbits with total spin $S$ a
good quantum number, the embedding algebra for EGUE(2)-$\cs$ and also for
EGOE(2)-$\cs$ is   $U(2\Omega) \supset U(\Omega) \otimes SU(2)$
\cite{Ko-06a,Ko-07}.  In particular the  EGOE(2)-$\cs$ with its extension
including mean-field one body part has been extensively used in the study
of quantum dots, small metallic grains and atomic nuclei
\cite{Ma-09,Ja-01,Pa-02}.

Wigner introduced in 1937 \cite{Wi-37} the spin-isospin $SU(4)$
supermultiplet scheme for nuclei. There is good evidence for the goodness
of this symmetry in some parts of the periodic table \cite{Pa-78} and also
more recently there is new interest in $SU(4)$ symmetry for heavy N $\sim$
Z nuclei \cite{Va-00}. Therefore it is clearly of importance to define and
study embedded Gaussian unitary ensemble of random matrices generated by
random  two-body interactions with $SU(4)$ symmetry, hereafter called
EGUE(2)-$SU(4)$. Given $m$ fermions (nucleons) in $\Omega$ number of sp
levels with spin  and isospin degrees of freedom, for $SU(4)$ scalar 
Hamiltonians, the symmetry algebra is $U(4\Omega)\supset  U(\Omega)\otimes
SU(4)$ and all the states within an $SU(4)$ irrep will be degenerate in
energy. Our purpose in this paper is to define EGUE(2)-$SU(4)$,  develop
analytical formulation for solving the ensemble and report the first
results for lower order cross correlations generated by this ensemble. Now
we will give a preview.

Section 2 gives a brief discussion of  $SU(4)$ algebra. EGUE(2)-$SU(4)$
ensemble is defined in Section 3. Also given here is the mathematical
formulation based on Wigner-Racah algebra of the embedding $U(4\Omega)
\supset U(\Omega) \otimes SU(4)$ algebra for solving the ensemble. In
Section 4, analytical formulas for $m$ fermion $U(\Omega)$ irreps
$f_m=\{4^r,p\}$ are given for the covariances in
energy centroids and spectral variances generated by this ensemble.
Section 5 gives discussion of some numerical results. Finally Section 6
gives summary and future outlook.

\section{Preliminaries of $U(4\Omega)\supset U(\Omega)\otimes SU(4)$
algebra}

Let us begin with $m$ nucleons distributed in $\Omega$ number of orbits
each with spin ($\cs=\spin$) and isospin ($\ct=\spin$)  degrees of freedom.
Then the total number of sp states is $N=4\Omega$ and the spectrum
generating  algebra is $U(4\Omega)$ . The sp states in uncoupled
representation are $a^\dagger_{i,\alpha} \l|0\ran = \l|i,\alpha\ran$  with
$i=1,2,\ldots,\Omega$ denoting the spatial orbits and $\alpha=1,2,3,4$ are
the four spin-isospin states $\l|m_\cs\,,m_\ct\ran=\l|\spin,\spin\ran$,
$\l|\spin,-\spin\ran$,  $\l|-\spin,\spin\ran$ and  $\l|-\spin,-\spin\ran$
respectively. The $(4\Omega)^2$  number of operators $C_{i\alpha;j\beta}$
generate $U(4\Omega)$ algebra. For $m$ fermions, all states belong to the
$U(4\Omega)$ irrep $\{1^m\}$. In uncoupled  notation,  $C_{i\alpha;j\beta}
= a^\dagger_{i,\alpha}a_{j,\beta}$.  Similarly $U(\Omega)$ and $U(4)$
algebras are generated by $A_{ij}$ and $B_{\alpha\beta}$ respectively,
where  $A_{ij} = \sum_{\alpha=1}^{4}\; C_{i\alpha;j\alpha}$ and
$B_{\alpha\beta} = \sum_{i=1}^{\Omega} \; C_{i\alpha;i\beta}$. The number
operator $\hat{n}$, the spin operator $\hat{S}= S_\mu^1$, the isospin
operator $\hat{T}=T_\mu^1$ and the Gamow-Teller operator
$\sigma\tau=(\sigma\tau)_{\mu,\mu^\pr}^{1,1}$ of $U(4)$ in spin-isospin
coupled notation are \cite{Ko-06},
\begin{eqnarray}
&& \hat{n} = 2\dis\sum_i \caa_{ii;0,0}^{0,0}\;,\;\;\;\;
S^1_\mu = \dis\sum_i \caa_{ii;\mu,0}^{1,0}\;,\;\;\;
T^1_\mu = \dis\sum_i \caa_{ii;0,\mu}^{0,1}\;, \nonumber \\
&& (\sigma\tau)_{\mu,\mu^\pr}^{1,1} = \dis\sum_i \caa_{ii;\mu,\mu^\pr}
^{1,1}\;;\;\;\;\;
\caa_{ij;\mu_\cs,\mu_\ct}^{s,t} = \l(a^\dagger_{i}\tilde{a}_{j}\r)_
{\mu_\cs,\mu_\ct}^{s,t}\;.
\label{eq.6}
\end{eqnarray}
Note that $\tilde{a}_{j;\mu_\cs,\mu_\ct} = (-1)^{1+\mu_\cs +\mu_\ct}
a_{j;-\mu_\cs,-\mu_\ct}$. These 16 operators form $U(4)$ algebra. Dropping
the number operator, we have $SU(4)$ algebra.

For the $U(4)$ algebra, the irreps are characterized by the partitions
$\{F\}=\{F_1,F_2,F_3,F_4\}$ with $F_1\geq F_2\geq F_3\geq F_4 \geq 0$ and
$m=\sum_{i=1}^4 F_i$.  Note that $F_\alpha$ are the eigenvalues of
$B_{\alpha\alpha}$. Due to the antisymmetry constraint on the total
wavefunction, the orbital space $U(\Omega)$ irreps $\{f\}$ are given by
$\{f\}=\{\widetilde{F}\}$ which is obtained by changing rows to columns in
$\{F\}$. It is important to note that, due to this symmetry constraint, for
the irrep $\{F\}$ each  $F_j\leq \Omega$ where $j=1,2,3,4$ and for the
irrep $\{f\}$ each  $f_i\leq 4$ with $i=1,2,\ldots,\Omega$.  The irreps for
the $SU(4)$ group are characterized by three rowed Young shapes
$\{F^\pr\}=\{F^\pr_1,F^\pr_2,F^\pr_3\}=\{ F_1-F_4,F_2-F_4,F_3-F_4\}$. Also
they can be mapped to $SO(6)$ irreps $[P_1,P_2,P_3]$ as the $SU(4)$ and
$SO(6)$ algebras are isomorphic to each other,  $[P_1,P_2,P_3] =
[(F_1+F_2-F_3-F_4)/2,\,(F_1-F_2+F_3-F_4)/2,\, (F_1-F_2-F_3+F_4)/2]$. Before
proceeding further, let us examine the quadratic Casimir invariants of
$U(\Omega)$, $U(4)$, $SU(4)$ and $SO(6)$ algebras. For example,
\beq
&& C_2\l[U(\Omega)\r] = \dis\sum_{i,j} A_{ij}A_{ji}
= \hat{n}\Omega - \dis\sum_{i,j,\alpha,\beta}
a^\dagger_{i,\alpha}a^\dagger_{j,\beta}a_{j,\alpha}a_{i,\beta}\;,
\nonumber \\
&& C_2\l[ U(4)\r] = \dis\sum_{\alpha,\beta} B_{\alpha,\beta}
B_{\beta,\alpha}
\Rightarrow C_2\l[U(\Omega)\r] + C_2\l[ U(4)\r] = \hat{n}\l(\Omega + 4\r)
\;.
\label{eq.12}
\eeq
Also, in terms of spin, isospin and Gamow-Teller operators, $C_2\l[
SU(4)\r] = C_2\l[ SO(6)\r] = S^2 + T^2 +  (\sigma \tau)\cdot (\sigma
\tau)$. Now we have the general results,
\beq
&& \lan C_2\l[ U(4)\r] \ran^{\{F\}} = \dis\sum_{i=1}^4 F_i(F_i+5-2i) =
\lan C_2\l[ SU(4)\r] + \dis\frac{\hat{n}^2}{4} \ran^{\{F^\pr\}}\;,
 \nonumber \\
&& \lan C_2\l[ SO(6)\r] \ran^{\l[P\r]} = \lan C_2\l[ SU(4)\r]\ran^
{\{F^\pr\}} = P_1(P_1+4) + P_2(P_2+2) + P_3^2\;.
\label{eq.17}
\eeq
In order to understand the significance of $SU(4)$ symmetry, let us
consider  the space exchange or the Majorana operator $M$ that exchanges
the spatial  coordinates of the particles and leaves the spin-isospin
quantum numbers unchanged,
\be
M \l| i,\alpha,\alpha^\pr ; j,\beta,\beta^\pr \ran =
\l| j,\alpha,\alpha^\pr ; i,\beta,\beta^\pr \ran\;,
\label{eq.18}
\ee
where $\alpha,\beta$ are labels for spin and $\alpha^\pr,\beta^\pr$ are
labels for isospin. As $\l| i,\alpha,\alpha^\pr ; j,\beta,\beta^\pr \ran
=a^\dagger_{i,\alpha,\alpha^\pr}a^\dagger_{j,\beta,\beta^\pr} \l| 0 \ran$,
Eqs. (\ref{eq.18}), (\ref{eq.12}) and (\ref{eq.17}) in that order will give,
\beq
2\;M & = & \dis\sum_{i,j,\alpha,\beta,\alpha^\pr,\beta^\pr}
\l( a^\dagger_{j,\alpha,\alpha^\pr} a^\dagger_{i,\beta,\beta^\pr}\r)
\l( a^\dagger_{i,\alpha,\alpha^\pr} a^\dagger_{j,\beta,\beta^\pr}\r)^
\dagger  \nonumber \\
& = & C_2\l[U(\Omega)\r] - \Omega \hat{n} = 4 \hat{n} - C_2\l[ U(4)\r]
\nonumber \\
\Rightarrow \alpha\,M & = & \alpha\l\{2\hat{n} \l( 1+\dis\frac{\hat{n}}
{16}\r) - \spin C_2\l[ SU(4)\r] \r\}\;.
\label{eq.21}
\eeq
The preferred $U(\Omega)$ irrep for the ground state of a $m$ nucleon system
is the most symmetric one. Therefore $\lan C_2\l[U(\Omega) \r]
\ran$  should be maximum for the ground state irrep. This implies,
as seen from Eq. (\ref{eq.21}),  the strength $\alpha$ of $M$ must be
negative.  As a consequence, as follows from the last equality in Eq.
(\ref{eq.21}),  the ground states are labeled by $SU(4)$ irreps with
smallest eigenvalue for the quadratic Casimir invariant  consistent with a
given ($m,T_z$), $T=|T_z|$.

For even-even nuclei, the ground state $SO(6)$ irreps are  $T=|$N$-$Z$|/2$.
For odd-odd N=Z nuclei, the ground state is  $\l[1\r]$ and for N $\neq$ Z
nuclei, it is $\l[ T,1\r]$. For odd-A nuclei, the irreps are $\l[
T,\spin,\pm\spin\r]$. Therefore, for N=Z even-even, N=Z odd-odd and N=Z$\pm
1$ odd-A nuclei the $U(\Omega)$ irreps for the ground states are
$\{4^r\}$, $\{4^r,2\}$, $\{4^r,1\}$ and $\{4^r,3\}$ with spin-isospin
structure being $(0,0)$, $(1,0) \oplus (0,1)$, $(\spin,\spin)$, and
$(\spin,\spin)$ respectively. For simplicity, in this paper we will present
final results only for these $U(\Omega)$ irreps. Other irreps will be
considered elsewhere. Now we will define EGUE(2)-$SU(4)$ ensemble and
derive some of its properties.

\begin{figure}
\epsfxsize=3in
\centerline{\epsfbox{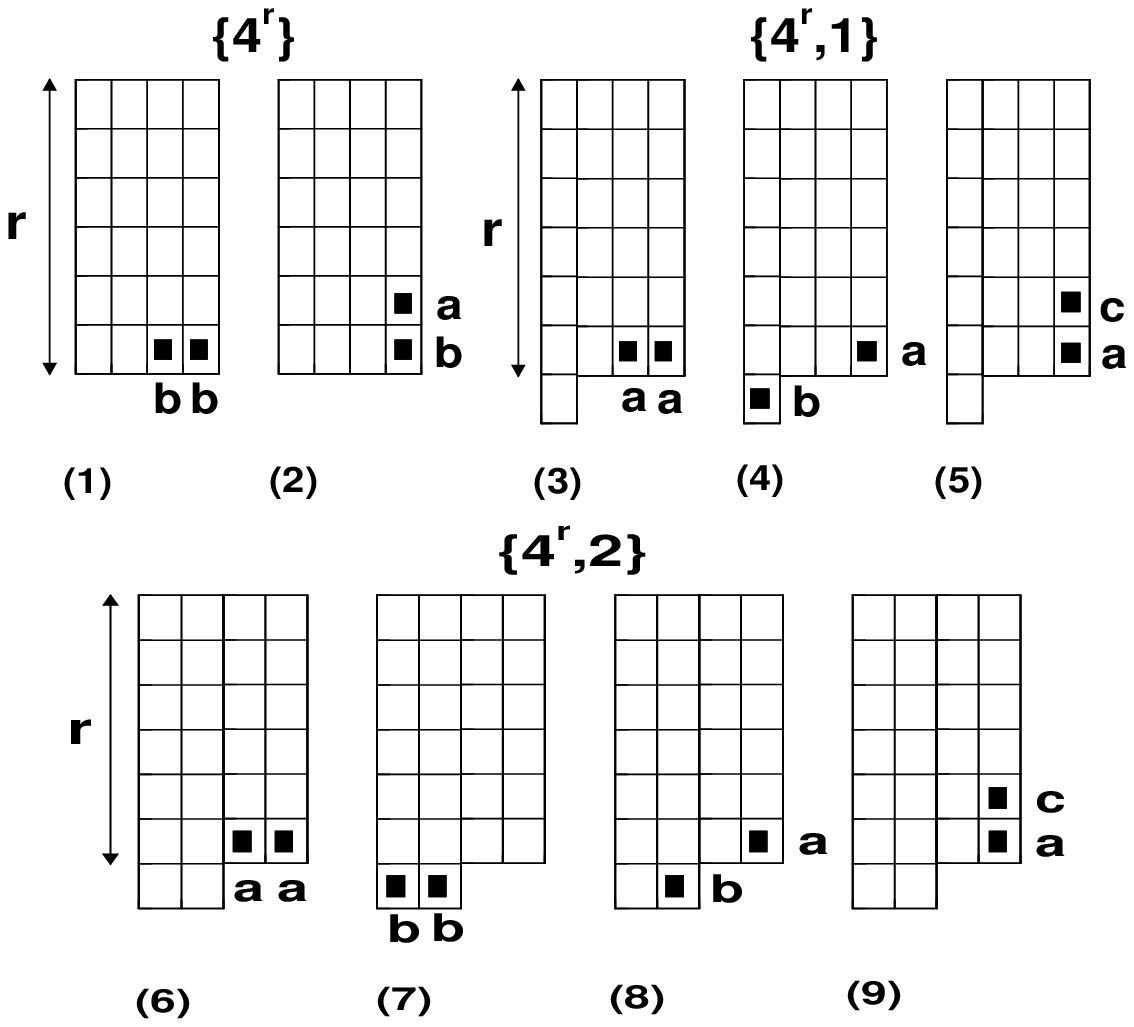}}
\noindent {\bf Figure 1.} Schematic representation of the Young tableaux
$f_m=\{4^r,p\}$
with $p=0$, $1$, and $2$. (1) $f_m=\{4^r\}$, $f_2=\{2\}$, $f_{m-2}=
\{4^{r-1},2\}$ and here $a=b$; (2) $f_m=\{4^r\}$, $f_2=\{1^2\}$,
$f_{m-2}= \{4^{r-2},3^2\}$ and $\tau_{ab}=1$; (3) $f_m=\{4^r,1\}$,
$f_2=\{2\}$, $f_{m-2}=\{4^{r-1},2,1\}$ and here $a=b$; (4) $f_m=\{4^r,1\}$,
$f_2=\{2\},\{1^2\}$, $f_{m-2}=\{4^{r-1},3\}$ and $\tau_{ab}=4$; (5)
$f_m=\{4^r,1\}$, $f_2=\{1^2\}$, $f_{m-2}=\{4^{r-2},3^2,1\}$ and
$\tau_{ac}=-1$; (6) $f_m=\{4^r,2\}$, $f_2=\{2\}$, $f_{m-2}=
\{4^{r-1},2^2\}$ and here $a=b$; (7) $f_m=\{4^r,2\}$, $f_2=\{2\}$,
$f_{m-2}=\{4^r\}$ and here $a=b$; (8) $f_m=\{4^r,2\}$, $f_2=\{2\},\{1^2\}$,
$f_{m-2}= \{4^{r-1},3,1\}$, $\tau_{ab}=3$; (9) $f_m=\{4^r,2\}$,
$f_2=\{1^2\}$, $f_{m-2}=\{4^{r-2},3^2,2\}$ and $\tau_{ac}=-1$.
\end{figure}

\section{EGUE(2)-$SU(4)$ ensemble: Definition and $U(4\Omega) \supset
U(\Omega)  \otimes SU(4)$ Wigner-Racah algebra for covariances}

Here we follow closely the approach used for EGUE(2)-$\cs$ recently
\cite{Ko-07}. Let us begin with normalized two-particle states $\l.\l|f_2
F_2; v_2 \beta_2\r.\ran$ where the $U(4)$ irreps  $F_2=\{1^2\}$  and
$\{2\}$  and the corresponding $U(\Omega)$ irreps $f_2$ are $\{2\}$
(symmetric) and $\{1^2\}$ (antisymmetric) respectively.  Similarly $v_2$
are additional quantum numbers that belong to $f_2$ and $\beta_2$ belong to
$F_2$. As $f_2$ uniquely defines $F_2$, from now on we will drop $F_2$
unless it is explicitly needed and also we will use the $f_2
\leftrightarrow F_2$ equivalence whenever needed.  With $A^\dagger(f_2 v_2
\beta_2)$ and $A(f_2 v_2 \beta_2)$ denoting creation and annihilation
operators for the normalized two particle states, a general two-body
Hamiltonian $H$ preserving $SU(4)$ symmetry can be written as
\be
H=\dis\sum_{f_2, v_2^i, v_2^f, \beta_2}\;V_{f_2 v_2^i v_2^f}(2)\;
A^\dagger(f_2 v^f_2 \beta_2)\,A(f_2 v^i_2 \beta_2)\;.
\label{eq.22}
\ee
In Eq. (\ref{eq.22}), $V_{f_2 v_2^i v_2^f}(2)= \lan f_2 v^f_2 \beta_2 \mid H
\mid f_2 v^i_2 \beta_2\ran$ independent of the $\beta_2$'s.  For
EGUE(2)-$SU(4)$ the $V_{f_2 v_2^i v_2^f}(2)$'s are independent Gaussian 
variables with zero center and variance given by (with bar representing 
ensemble average),
\be
\overline{V_{f_2 v_2^1 v_2^2}(2)\;V_{f_2^\pr v_2^3 v_2^4}(2)}
= \delta_{f_2 f_2^\pr} \delta_{v_2^1 v_2^4} \delta_{v_2^2 v_2^3}\,
(\lambda_{f_2})^2\;.
\label{eq.23}
\ee
Thus $V(2)$ is a direct sum of GUE matrices for $F_2=\{2\}$ and
$F_2=\{1^2\}$ with variances $(\lambda_{f_2})^2$ for the diagonal matrix
elements and  $(\lambda_{f_2})^2/2$ for the real and imaginary parts of the
off-diagonal matrix elements. As discussed before for EGUE(k) \cite{Ko-05}
and EGUE(2)-$\cs$ \cite{Ko-07}, tensorial decomposition of $H$ with respect
to the embedding algebra $U(\Omega) \otimes SU(4)$ plays a crucial role in
generating analytical  results; as in \cite{Ko-05}, the $U(\Omega)
\leftrightarrow SU(\Omega)$ correspondence is used throughout and therefore
we use $U(\Omega)$ and  $SU(\Omega)$ interchangeably. As $H$ preserves
$SU(4)$, it is a scalar  in the $SU(4)$ space. However with respect to
$SU(\Omega)$, the tensorial characters, in Young tableaux notation, for
$f_2=\{2\}$ are $\bF_\nu=\{0\}$, $\{21^{\Omega-2}\}$ and
$\{42^{\Omega-2}\}$ with $\nu=0,1$ and 2 respectively. Similarly for
$f_2=\{1^2\}$ they are $\bF_\nu= \{0\}$, $\{21^{\Omega-2}\}$ and $\{2^2
1^{\Omega-4}\}$ with $\nu=0,1,2$ respectively. Note that $\bF_\nu=f_2
\times \overline{f_2}$ where $\overline{f_2}$ is the irrep conjugate to
$f_2$ and the $\times$ denotes Kronecker product. Then we can define
unitary tensors $B$'s that are scalars  in $SU(4)$ space,
\beq
B(f_2 \bF_\nu \omega_\nu) & = & \dis\sum_{v_2^i,v_2^f, \beta_2}\,
A^\dagger(f_2 v^f_2 \beta_2)\, A(f_2 v^i_2 \beta_2)\, \lan f_2
v_2^f\;\overline{f_2}\,\overline{v_2^i} \mid \bF_\nu \omega_\nu\ran
\nonumber \\
&\times&
\lan F_2 \beta_2\;\overline{F_2}\,\overline{\beta_2} \mid 0 0\ran\;.
\label{eq.24}
\eeq
In Eq. (\ref{eq.24}), $\lan f_2 --- \ran$ are $SU(\Omega)$  Wigner
coefficients  and $\lan F_2 --- \ran$ are $SU(4)$ Wigner coefficients.
The expansion of $H$ in terms of $B$'s is,
\be
H=\dis\sum_{f_2, \bF_\nu, \;\omega_\nu}\;W(f_2 \bF_\nu \omega_\nu)\,
B(f_2 \bF_\nu \omega_\nu)\;.
\label{eq.25}
\ee
The expansion coefficients $W$'s follow from the orthogonality of the
tensors $B$'s with respect to the traces over fixed $f_2$ spaces. Then we
have the most important relation needed for all the results given ahead,
\be
\overline{W(f_2 \bF_\nu \omega_\nu)W(f^\pr_2 \bF^\pr_\nu \omega^\pr_\nu)}
=\delta_{f_2 f^\pr_2} \delta_{\bF_\nu \bF^\pr_\nu} \delta_{\omega_\nu
\omega^\pr_\nu} \, (\lambda_{f_2})^2 d(F_2)\,.
\label{eq.26}
\ee
This is derived starting with Eq. (\ref{eq.23}) and substituting,  in two
particle matrix elements $V$, for $H$ the expansion given by Eq.
(\ref{eq.25}). Also used are the sum rules for Wigner coefficients
appearing in Eq. (\ref{eq.24}).
\begin{figure}[ht]
\epsfxsize 4in
\centerline{\epsfbox{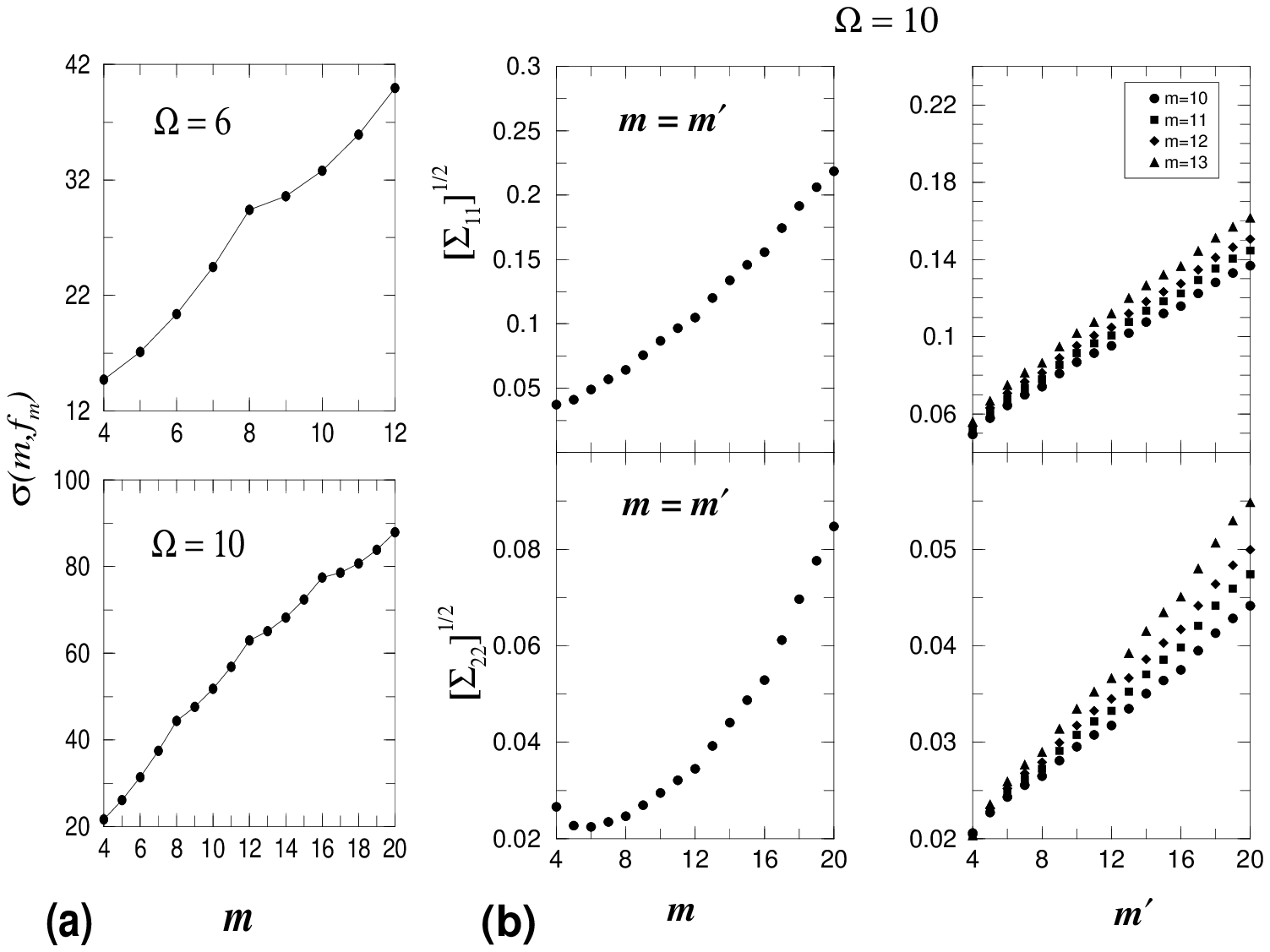}}
\noindent{\bf Figure 2.} (a) Widths $\sigma(m,f_m)=[\;\overline{\lan
H^2\ran^{m,f_m}}\;] ^{1/2}$ for $\Omega=6$ and $\Omega=10$ examples.  (b)
Cross correlations for $\Omega=10$ examples. Note that $f_m=\{4^r,p\}$; 
$m=4r+p$ and $f_{m^\pr}=\{4^s,q\}$; $m^\pr=4s+q$.  See text for
details.
\end{figure}
\begin{center}
{\bf Table 1.} $P^{f_2}(m,f_m)$ for $f_m=\{4^r,p\}$; $p=0,1,2$ and $3$ and
$\{f_2\}=\{2\},\{1^2\}$. 
{\footnotesize{
\begin{tabular}{ccc}
\hline \\
 & $\;\;\;\;\;\;\;\;\;\;\;\;\;\;\;\;\;\;\;\;\;\;\;\;\;\;\;\;\;\;\;\;\;
\;\;\;\;\;\;\;\;P^{f_2}(m,f_m)$ & \\ \\
$f_m$ & $f_2=\{2\}$ & $f_2=\{1^2\}$ \\ \\
\hline \\
$\{4^r\}$ & $-3r(r+1)$ & $-5r(r-1)$ \\ \\
$\{4^r,1\}$ & $-\dis\frac{3r}{2}(2r+3)$ & $-\dis\frac{5r}{2}(2r-1)$\\ \\
$\{4^r,2\}$ & $-(3r^2+6r+1)$ & $-5r^2$ \\ \\
$\{4^r,3\}$ & $-\dis\frac{3}{2}(r+2)(2r+1)$ & $-\dis\frac{5r}{2}(2r+1)$\\ \\
\hline
\end{tabular}}}
\end{center}

Turning to $m$ particle $H$ matrix elements, first we denote the
$U(\Omega)$ and $U(4)$ irreps by $f_m$ and $F_m$ respectively.
Correlations generated by EGUE(2)-$SU(4)$ between states with $(m,f_m)$ and
$(m^\pr,f_{m^\pr})$ follow from the covariance between the $m$-particle
matrix elements of $H$. Now using Eqs. (\ref{eq.25}) and (\ref{eq.26})
along with the Wigner-Eckart theorem applied  using $SU(\Omega) \otimes
SU(4)$ Wigner-Racah algebra (see for example \cite{Dr-74}) will give
\beq
&& \overline{H_{f_m v_m^i v_m^f}\,H_{f_{m^\pr} v_{m^\pr}^i v_{m^\pr}^f}}
\nonumber \\ \nonumber
&& =
\overline{\lan f_m F_m v_m^f \beta \mid H \mid f_m F_m v_m^i \beta\ran
\lan f_{m^\pr} F_{m^\pr} v_{m^\pr}^f \beta^\pr \mid H \mid f_{m^\pr}
F_{m^\pr} v_{m^\pr}^i \beta^\pr \ran } \nonumber
\\ \nonumber \\
&& = \dis\sum_{f_2, \bF_\nu,\; \omega_\nu} \; \dis\frac{(\lambda_{f_2})^2}
{d(f_2)}\;
\dis\sum_{\rho,\rho^\pr}\; \lan f_m \mid\mid\mid B(f_2 \bF_\nu)
\mid\mid\mid
f_m\ran_\rho\;
\lan f_{m^\pr} \mid\mid\mid B(f_2 \bF_\nu) \mid\mid\mid f_{m^\pr}
\ran_{\rho^\pr} \nonumber \\
&& \times \lan f_m v_m^i\;\bF_\nu \omega_\nu
\mid f_m v_m^f\ran_\rho\;
\lan f_{m^\pr} v_{m^\pr}^i\;\bF_\nu \omega_\nu \mid f_{m^\pr}
v_{m^\pr}^f\ran_{\rho^\pr}\,;
\nonumber \\
&& \lan f_m \mid\mid\mid
B(f_2 \bF_\nu) \mid\mid\mid f_m\ran_\rho\,=
\dis\sum_{f_{m-2}}\;F(m)\, \dis\frac{\cn_{f_{m-2}}}{\cn_{f_m}} \;
\dis\frac{U(f_m \overline{f_2} f_m f_2; f_{m-2} \bF_\nu)_\rho}{U(f_m
\overline{f_2} f_m f_2; f_{m-2} \{0\})} \;.
\label{eq.29}
\eeq
Here the summation in the last equality is over the multiplicity index
$\rho$ and this arises  as  $f_m \otimes \bF_\nu$ gives in general more
than once the irrep $f_m$. In Eq. (\ref{eq.29}), $F(m)=-m(m-1)/2$, $d(f_m)$
is dimension with respect to $U(\Omega)$ and $\cn_{f_m}$ is dimension with
respect to the  $S_m$ group; formulas for these dimensions are given in
\cite{Wy-70}. Similarly, $\lan \ldots \ran$ and $U(\ldots)$ are
$SU(\Omega)$ Wigner and Racah coefficients respectively.

\section{Lower order cross correlations in EGUE(2)-$SU(4)$}

Lower order cross correlations between states with different $(m,f_m)$ are
given by the  normalized bivariate moments $\Sigma_{rr}\l(m,f_m: m^\pr
,f_{m^\pr}\r)$, $r=1,2$ of the two-point function $S^\rho$ where, 
with $\rho^{m,f_m}(E)$ defining fixed-$(m,f_m)$ density of states,
\beq
&& S^{m f_m:m^\pr f_{m^\pr}}(E,E^\pr) = \overline{\rho^{m,f_m}(E)
\rho^{m^\pr ,
f_{m^\pr}}(E^\pr)} - \overline{\rho^{m,f_m}(E)}\;\;\overline{\rho^{m^\pr ,
f_{m^\pr}}(E^\pr)}\;\;; \nonumber\\ \nonumber \\
&& \Sigma_{11}\l( {m,f_m}: {m^\pr ,f_{m^\pr}}\r) = \overline{
\lan H \ran^{m,f_m} \; \lan H \ran^{m^\pr,f_{m^\pr}}}/\dis\sqrt{\,
\overline{\lan H^2 \ran^{m,f_m}} \; \overline{\lan H^2 \ran^{m^\pr,
f_{m^\pr}}}}\;, \nonumber\\ \nonumber \\
&& {\Sigma}_{22}\l( {m,f_m}: {m^\pr ,f_{m^\pr}}\r) = \overline{
\lan H^2 \ran^{m,f_m} \; \lan H^2 \ran^{m^\pr,f_{m^\pr}}}/\l[\,
\overline{\lan H^2 \ran^{m,f_m}} \; \overline{\lan H^2 \ran^{m^\pr,
f_{m^\pr}}}\,\r] -1\;.
\label{eq.30}
\eeq
In Eq. (\ref{eq.30}), $\overline{\lan H^2 \ran^{m,f_m}}$ is the second
moment (or variance) of $\overline{\rho^{m,f_m}(E)}$ and  its centroid
$\overline{\lan H\ran^{m,f_m}}=0$ by definition. We begin with
$\overline{\lan H \ran^{m,f_m} \; \lan H \ran^{m^\pr,f_{m^\pr}}}$. As $\lan
H \ran^{m,f_m}$ is the trace of $H$  (divided by dimensionality)  in
$(m,f_m)$ space, only $\bF_\nu = \{0\}$ will generate this. Then trivially,
\be
\overline{\lan H \ran^{m,f_m} \; \lan H \ran^{m',f_{m^\pr}}}
 = \dis\sum_{f_2} \dis\frac{\l(\lambda_{f_2}\r)^2}{d(f_2)}\;P^{f_2}
(m,f_m)\;P^{f_2}(m^\pr,f_{m^\pr})\;. 
\label{eq.31} 
\ee 
Note that
$P^{f_2}(m,f_m)=F(m)\sum _{f_{m-2}}\;[\cn_{f_{m-2}}/\cn_{f_m}]$. 
The formulas for $P^{f_2}(m,f_m)$ are given in Table 1.
Writing $\overline{\lan H^2
\ran^{m,f_m}}$ explicitly in terms  of $m$ particle $H$ matrix
elements, $\overline{\lan H^2 \ran^{m,f_m}} =
[d(f_m)]^{-1}\sum_{v_m^1 ,  v_m^2} \, \overline{H_{f_m v_m^1
v_m^2}\,H_{f_m v_m^2 v_m^1}}$, and applying Eq. (\ref{eq.29}) and
the orthonormal properties of the $SU(\Omega)$ Wigner coefficients
lead to
\be
\overline{\lan H^2 \ran^{m,f_m}} = \dis\sum_{f_2}
\dis\frac{(\lambda_{f_2})^2}{d(f_2)}
\dis\sum_{\nu=0,1,2}\cq^\nu(f_2:m,f_m)\;.
\label{eq.35}
\ee

\begin{center}
{\bf Table 2.}  $\overline{\lan H^2\ran^{m,f_m}}$, 
$\cq^{\nu=1,2}(f_2:m,f_m)$ and $R^{\nu=1}(m,f_m)$ for some examples.
{\footnotesize{
\begin{tabular}{cccl}
\hline \\
$f_m$ & & & $\overline{\lan H^2\ran^{m,f_m}}$ \\ \\
\hline \\
$\{4^r\}$ & & & $\dis\frac{r(\Omega-r+4)}{2}\l[\lambda^2_{\{2\}}3(r+1)
(\Omega-r+3)+\lambda^2_{\{1^2\}} 5(r-1)(\Omega-r+5)\r]$ \\ \\
$\{4^r,1\}$ & & & $\dis\frac{r(\Omega-r+4)}{4}\l[  \lambda^2_{\{2\}}
\{ 6r(\Omega-r+1)+9\Omega+15\}\r.$ \\
& & & $\l.+ \lambda^2_{\{1^2\}}
5\{ 2r(\Omega-r+5)-\Omega-9\}\r]$ \\ \\
$\{4^r,2\}$ & & & $\lambda^2_{\{2\}}\spin
\l[ 3r^4-6(\Omega+2)r^3+(3\Omega^2+6\Omega-5)r^2\r.$ \\
& & & $\l.+(\Omega+2)(6\Omega+17)r+\Omega(\Omega+1)\r]$ \\ 
& & & $+\lambda^2_{\{1^2\}} \dis\frac{5r}{2}(\Omega-r+4)\{
(\Omega+4)r-r^2-3\}$ \\ \\
$\{4^r,3\}$ & & & $\dis\frac{1}{4}\l[\lambda^2_{\{2\}}3(r+2)(\Omega-r+2)
(2r\Omega-2r^2+6r+\Omega+1)\r.$ \\
& & & $\l.+\lambda^2_{\{1^2\}}5r(\Omega-r+4)
(2r\Omega-2r^2+6r+\Omega-1)\r]$ \\ \\
\hline \\
$f_m$ & $f_2$ & $\nu$ & $Q^\nu(f_2:m,f_m)$ \\ \\
\hline \\
$\{4^r\}$ & $\{2\}$ & $1$ & $\dis\frac{9r(r+1)^2(\Omega-r)(\Omega+1)
 (\Omega+4)}{2(\Omega+2)}$ \\
 & & $2$ & $\dis\frac{3r\Omega(r+1)(\Omega-r+1)(\Omega-r)
 (\Omega+4) (\Omega+5)}{4(\Omega+2)}$ \\
 & $\{1^2\}$ & $1$ & $\dis\frac{25r(r-1)^2(\Omega-r)(\Omega-1)(\Omega+4)}
 {2(\Omega-2)}$ \\
 & & $2$ & $\dis\frac{5r\Omega(r-1)(\Omega+3)(\Omega+4)(\Omega-r)
 (\Omega-r-1)}{4(\Omega-2)}$ \\
\hline \\ 
$f_m$ & & & $R^{\nu=1}(m,f_m)$  \\ \\
\hline \\ \\
$\{4^r\}$ & & &
$-\dis\frac{15r}{2}\dis\sqrt{\dis\frac{\Omega^2-1}{\Omega^2-4}}
(r^2-1)(\Omega-r)(\Omega+4)$ \\ \\
\hline
\end{tabular}
}}
\end{center}

\noindent The functions $\cq^\nu(f_2:m,f_m)$ involve $SU(\Omega)$ Racah
coefficients and they are available in various tables in a complex
form involving functions of $\tau_{ab}$ \cite{He-74}. Here
$\tau_{ab}$ are the axial distances for a given Young tableaux (see
Fig. 1 for examples). Evaluating all the functions, we have derived
analytical formulas for $\cq^\nu(f_2:m,f_m)$ and also for 
$\overline{\lan H^2 \ran^{m,f_m}}$. Some of these results are given in 
Table 2.
It is easily seen that $\cq^{\nu=0}(f_2:m,f_m) =
\l[P^{f_2}(m,f_m) \r]^2$. Results in Tables 1 and 2 will
give formulas for the covariances ${\Sigma}_{11}$ in energy
centroids.  Similarly, analytical results for covariances
${\Sigma}_{22}$ in  spectral variances are derived using Eqs.
(\ref{eq.29}) and (\ref{eq.30}) and then, 
\beq 
&& {\Sigma}_{22}(m,f_m;m^\pr,f_{m^\pr}) = \dis\frac{X_{\{2\}}+
X_{\{1^2\}} +4X_{\{1^2\}\{2\}}}{\overline{\lan H^2\ran^{m,f_m}}\;
\overline{\lan H^2\ran^{m^\pr,f_{m^\pr}}}}\;;\nonumber \\
&& X_{f_2} = \dis\frac{2(\lambda_{f_2})^4}{\l[d(f_2)\r]^2}
\dis\sum_{\nu=0,1,2} \l[ d(\bF_\nu)\r]^{-1}\cq^\nu(f_2:m,f_m)
\cq^\nu(f_2:m^\pr,f_{m^\pr})\;,
\nonumber \\
&& X_{\{1^2\}\{2\}} =
\dis\frac{\lambda^2_{\{2\}}\lambda^2_{\{1^2\}}}{d(\{2\})d(\{1^2\})}
\dis\sum_{\nu=0,1} \l[ d(\bF_\nu)\r]^{-1} R^\nu(m,f_m)\;
R^\nu(m^\pr,f_{m^\pr})\;. 
\label{eq.37}
\eeq
Here $d(\bF_\nu)$ are dimension of the irrep $\bF_\nu$, and we have
$d(\{0\})=1$, $d(\{2,1^{\Omega-2}\})=\Omega^2-1$, $d(\{4,2^{\Omega-2}\})=
\Omega^2(\Omega+3)(\Omega-1)/4$, and $d(\{2^2,1^{\Omega-4}\})
=\Omega^2(\Omega-3)(\Omega+1)/4$. Again the functions $R^\nu$ involve
$SU(\Omega)$ Racah coefficients and $R^{\nu=0}(m,f_m) = P^{\{2\}}
(m,f_m)P^{\{1^2\}}(m,f_m)$. Formulas for $R^{\nu=1}(m,f_m)$   for
$f_m=\{4^r,p\}$, $p=0-3$ are derived and the result for $\{4^r\}$ is given
in Table 2 as an example. Complete tabulations for $\cq^{\nu=1,2}$ and
$R^{\nu=1}$ will be reported elsewhere. Equations (\ref{eq.31}), (\ref{eq.35}) 
and (\ref{eq.37}) are similar in structure to the corresponding equations
for EGUE(2)-$\cs$ \cite{Ko-07}.  However the functions $P$'s, $\cq$'s and
$R$'s are more complicated for EGUE(2)-$SU(4)$. 

\section{Results and discussion}

Numerical calculations are carried out for $\overline {\lan H^2 \ran^
{m,f_m}}$, $\Sigma_{11}$ and $\Sigma_{22}$ for some $\Omega=6$
[($2s1d$)-shell] and  $\Omega=10$ [($2p1f$)-shell] examples. Here we have
employed $\lambda_{\{1^2\}}^2=\lambda_{\{2\}}^2=1$. Fig. 2a shows the
variation in the spectral widths $\sigma(m,f_m)= [\;\overline{\lan
H^2\ran^{m,f_m}}\;]^{1/2}$ with particle number $m$. Notice the peaks at
$m=4r$; $r=2,3,\ldots$. Except for this structure, there are no other
differences between $\{4^r\}$ and $\{4^r,2\}$ systems i.e. for ground
states of even-even and odd-odd N=Z nuclei. Results for the cross 
correlations $\Sigma_{11}$ and $\Sigma_{22}$ are shown in Fig. 2b. It
is seen that $[\Sigma_{11}]^{1/2}$ and $[\Sigma_{22}]^{1/2}$ increases
almost linearly with $m$. At $m=4r$, $r=2,3,\ldots$ there is a slight dip
in $[\Sigma_{11}]^{1/2}$ as well as in $[ \Sigma_{22}]^{1/2}$. For
$\Omega=6$ with $m=m^\pr$, $[\Sigma_{11}]^{1/2} \sim 10-28$\% and 
$[\Sigma_{22}]^{1/2} \sim 6-16$\% as $m$ changes from 4 to 12. Similarly
for $m \neq m^\pr$, $[\Sigma_{11}]^{1/2}\sim 10-24$\% and $[
\Sigma_{22}]^{1/2}\sim 6-12$\%. The values are somewhat smaller for
$\Omega=10$ (see Fig. 2b) which is in agreement with the results obtained
for EGOE(2) for spinless fermions and EGOE(2)-$\cs$. For further
understanding we compare, for fixed $N$, these covariances with those for
EGUE(2) and EGUE(2)-$\cs$. Using the analytical formulas given in
\cite{Ko-05} for EGUE(2), \cite{Ko-07} for EGUE(2)-$\cs$ and the present
paper for EGUE(2)-$SU(4)$, it is found that the magnitude of the
covariances in energy centroids and spectral variances increases with
increasing symmetry. For example, with $N=24$ [so that $\Omega=12$ for
EGUE(2)-$\cs$ and $\Omega=6$ for EGUE(2)-$SU(4)$] the results are as
follows. For $m = m^\pr = 6$ ($m = m^\pr = 8$) we have: (i)
$[\Sigma_{11}]^{1/2} = 0.017 (0.026)$ and $[\Sigma_{22}]^{1/2} = 0.006
(0.006)$ for EGUE(2); (ii) for EGUE(2)-$\cs$ with $S=S^\pr=0$,
$[\Sigma_{11}]^{1/2} = 0.043 (0.066)$ and $[\Sigma_{22}]^{1/2} = 0.017
(0.021)$; (iii) for EGUE(2)-$SU(4)$, $[\Sigma_{11}]^{1/2} = 0.124 (0.16)$
and $[\Sigma_{22}]^{1/2} = 0.069 (0.082)$. As fluctuations are growing with
increasing symmetry, it is plausible to conclude that symmetries play a
significant role in generating chaos. From a different perspective a
similar conclusion was reached in \cite{PW-05} by Papenbrock and 
Weidenm\"{u}ller. As they state: ``While the number of independent random
variables decreases drastically as we follow this sequence, the complexity
of the (fixed) matrices which support the random variables, increases even
more. In that sense, we can say that in the TBRE, chaos is largely due to
the existence of (an incomplete set of) symmetries.'' 

\section{Summary and future outlook}

In summary, we have introduced the embedded ensemble EGUE(2)-$SU(4)$ in
this paper  and our main emphasis has been in presenting analytical
results. Our study is restricted to $U(\Omega)$ irreps of the type
$\{4^r,p\}$, $p=0,1,2$ and $3$. Using Eqs. (13)-(15) and the formulas for
the functions $P$'s, $\overline{\lan H^2\ran^{m,f_m}}$, $\cq$'s and $R$'s
given in Tables 1 and 2, cross correlations in spectra with different
$(m,f_m)$ irreps are studied with results presented in Fig. 2 and Section
5. See \cite{Pa-07,Ko-07,Ko-06a} for further discussion on the significance
of cross correlations generated by embedded ensembles (they will vanish for
GE's). Elsewhere we will discuss the results for EGOE(2)-$SU(4)$ and in the
limit $\Omega \to \infty$ the results for these two ensembles are expected
to coincide except for a difference in scale factors. 

In future we also plan to investigate EGUE(2)-$SU(4)$ for general
$U(\Omega)$ irreps for any $m$ and this is indeed feasible with the
tabulations for sums of Racah coefficients given in \cite{He-74}. Then it is
possible to examine the extent to which EGUE(2)-$SU(4)$, i.e. random
interactions with $SU(4)$ symmetry, carry the properties of Majorana or the
$C_2[SU(4)]$ operator. This study is being carried out and the results will
be presented elsewhere. With this, it is possible to understand the role of
random interactions in generating the differences in the ground state
structure of even-even and odd-odd N=Z nuclei. See \cite{Kir-07} for a
numerical random  matrix study of N=Z nuclei. In addition, just as the
pairing correlations in EGOE(1+2)-$\cs$ have been investigated recently
\cite{Ma-09},  it is possible to consider $SU(\Omega) \supset SU(3)$, where
$SU(3)$ is Elliott's $SU(3)$ algebra \cite{El-58}, and examine rotational
collectivity with random interactions. To this end we plan to analyze in
future expectation values of the quadratic Casimir  invariant of $SU(3)$ or
equivalently that of quadrupole-quadrupole ($Q.Q$)  operator over the
EGOE(1+2)-$SU(4)$ ensemble.  Finally, going beyond EGUE(2)-$SU(4)$, it is
both interesting and possible (by extending and applying the $SU(4) \supset
SU_S(2) \otimes SU_T(2)$ Wigner-Racah
algebra developed by Hecht and Pang \cite{Pan-69}) to define and investigate,
analytically, the ensemble with full $SU(4)-ST$ symmetry. In principle, it
is also possible to construct  the $m$ particle $H$ matrix, which is $SU(4)$
or $SU(4)-ST$ scalar, on a computer and analyze its properties
numerically but this is for future.

\acknowledgements

The present study has grown out of the discussions one of the authors (VKBK)
has had with O. Bohigas and H.A. Weidenm\"{u}ller in the 2008 Shanghai
`Nuclear Structure Physics' meeting.

\ed